\documentclass[10pt]{iopart}
\usepackage{graphicx}

\usepackage{amssymb}

\begin{document}
\jl{1}
\title{The distribution of extremal points of Gaussian scalar fields}
\author{Georg Foltin}
\address{Institut f\"ur Theoretische
Physik III, Heinrich--Heine--Universit\"at D\"usseldorf,
Universit\"atsstrasse
1, D--40225 D\"usseldorf, Germany}
\begin{abstract}
We consider the signed density of the extremal points of (two-dimensional) scalar fields with a Gaussian distribution. We assign a positive unit charge to the maxima and minima of the function and a negative one  to its saddles. At first, we compute the average density for a field in half-space with Dirichlet boundary conditions. Then we calculate the charge-charge correlation function (without boundary). We apply the general results to random waves and random surfaces.
Furthermore, we find a generating functional for the two-point function. Its Legendre transform is the integral over the scalar curvature of a  four-dimensional Riemannian manifold. 
\end{abstract}

\section{Introduction}
In two-dimensional systems with orientational degrees of freedom, it is often fruitful to focus on the nodal points (zeros, defects), and to ignore other degrees of freedom. The distribution of the nodal points keeps information about the degree of (orientational) order within the system, and about the topology of the `substrate'.
Well known examples are defects in liquid crystal films, vortices in thin superconductors and superfluid films (for an overview, see \cite{Cha95}), or chaotic wave functions in quantum mechanics \cite{Ber78}, microwave billiards \cite{Seb99,Bar02}, and optical speckle pattern \cite{Bar81}.

Here we take a closer look at the distribution of the nodal points of a gradient field or likewise the extremal points (maxima, minima, saddles) of  a scalar field $\phi$ with a Gaussian distribution\footnote{The distribution of $\phi$ is solely characterized by the correlation function $\left<\phi(\bi{r})\phi'(\bi{r}')\right>$. Higher moments are calculated with the help of the Wick theorem.}. The locations of the extrema of $\phi$ are the points, where $\phi$ is stationary, i.e. the gradient $\nabla\phi$ vanishes. The distribution of the stationary points was discussed earlier in the context of random surfaces \cite{Car56,Lon57a,Lon57b} and of  excursion sets \cite{Adl00,Tay03}.

The extremal points carry a topological charge $q$, which is the sign of the Jacobian right at the extremum $q=\textrm{sign}(\det(\partial_i\partial_j\phi))=\pm 1$. In two dimensions, the positive extrema are the maxima and minima of the function $\phi$, the negative extrema are the saddle points. We will compute moments of the
\textit{signed} density
\begin{equation}
\rho(\bi{r})=\sum_\alpha q_\alpha\delta^2(\bi{r}-\bi{r}_\alpha)
\end{equation}
where $\bi{r}_\alpha$ are the positions of the extrema with sign (charge) $q_\alpha$.

From the experimental point of view, measuring the \textit{absolute} mean values of the density, $|\rho|$, is much easier than measuring the usually small density difference between the density of the positive points minus the density of the negative ones. The signed density, however, is the much more appealing object from the mathematical point of view.
The signed density is subject to a topological constraint, and its averages are differential geometrical objects, as shown in \cite{Fol03a}. 

We represent the signed density of the extremal points of a $d$-dimensional scalar field $\phi$ through \cite{Lon57a,Lon57b}
\begin{equation}
\rho(\bi{r})=\det\left(\partial_i\partial_j\phi(\bi{r})\right)\delta^d\left(\nabla\phi(\bi{r})\right).
\end{equation}
In order to calculate the average of the absolute density $\left<|\rho|\right>$, one needs the simultaneous distribution of the $d(d+1)/2$ random variables $\partial_i\partial_j\phi$ and of the $d$ variables $\nabla\phi$, where both sets are Gaussian random variables. Therefore, we need to know all moments $\left<\partial_i\partial_j\phi\partial_k\partial_l\phi\right>$,  $\left<\partial_i\partial_j\phi\partial_k\phi\right>$, and $\left<\partial_i\phi\partial_j\phi\right>$
at coinciding points. For the averaged \textit{signed} density, however, we need less information. According to \cite{Fol03a} it is sufficient to know the moments
\begin{equation}
\label{mt}
g_{ij}(\bi{r})=\left<\partial_i\phi(\bi{r})\partial_j\phi(\bi{r})\right>
\end{equation} 
(which usually depend on the position $\bi{r}$).
The mean density $\left<\rho\right>$ turns out to be proportional to the (total) curvature of the $d$-dimensional Riemannian manifold which is described by the above metric tensor $g_{ij}$.   In two dimensions, the  
mean density is proportional to the Gaussian curvature $K$ of that manifold (`surface') (times the invariant area element $\sqrt{\det g_{ij}}$)
\begin{equation}
2\pi\left<\rho(\bi{r})\right>=K\sqrt{\det g_{ij}}.
\end{equation}
Also, higher-order correlation functions fit into this scheme. The correlation function
$\left<\rho(\bi{r}_1)\ldots\rho(\bi{r}_f)\right>$ is the total curvature of a $f\times d$ dimensional manifold with metric tensor\footnote{We define a $f\times d$ dimensional Gaussian scalar function $\Phi(\bi{r}_1,\ldots,\bi{r}_f)=\phi(\bi{r}_1)+\ldots+\phi(\bi{r}_f)$. It is obviously $\rho_\Phi(\bi{r}_1,\ldots,\bi{r}_f)=\rho(\bi{r}_1)\times\ldots\times\rho(\bi{r}_f)$, where $\rho_\Phi$ is the density of extrema of the field $\Phi$. We obtain the metric tensor $\left<\partial_{i\alpha}\Phi\partial_{j\beta}\Phi\right>=\left<\partial_i\phi(\bi{r}_\alpha)\partial_j\phi(\bi{r}_\beta)\right>$.}   
\begin{equation}
\label{metrictensor}
g_{i\alpha,j\beta}=\left<\partial_i\phi(\bi{r}_\alpha)\partial_j\phi(\bi{r}_\beta)\right>,\,\, i,j=1\ldots d,\,\alpha,\beta=1\ldots f
\end{equation}
where a pair of a Greek and a Roman index is seen as a single composite index. 
The charge-charge correlation function for a Gaussian field in two dimensions is therefore the curvature of a particular four-dimensional Riemann manifold. We shall see, that the `Einstein'-action, i.e. the covariant integral over the scalar curvature $R$ of the four-dimensional manifold, plays an important role as a generating functional, relating the two-point function and the correlation function of the scalar field $\phi$. 
Before we discuss the two-point function, we present a simple example of a field $\phi$ with a nonzero mean density  $\left<\rho(\bi{r})\right>$, namely a system with a straight boundary.
The plots of the charge-charge density and of the charge density near a wall are somewhat similar, since the latter quantity might be seen as the correlation of a charge with its mirror charge. The general results are applied to random waves, and to thermally fluctuating surfaces (fluid membranes).

\section{Density of extremal points near a boundary}
\label{sec-bound}
We calculate the signed density of extremal points of a Gaussian distributed random function $\phi(x,y)$ in the upper half space $y>0$ and assume Dirichlet boundary conditions $\phi(x,y=0)=0$ (see \cite{Ber02}). We begin with a field $\phi(\bi{r})$ in Euclidean, two-dimensional space with an isotropic and translationally invariant correlation function
\begin{equation}
\left<\phi(x,y)\phi(x',y')\right>=G\left(\sqrt{(x-x')^2+(y-y')^2}\right).
\end{equation}
The correlation function for the upper half space is obtained by subtracting the appropriate mirror image
\begin{eqnarray}
\lefteqn{G_+(x,y|x',y')\equiv\left<\phi(x,y)\phi(x',y')\right>}\nonumber\\
&=&G\left(\sqrt{(x-x')^2+(y-y')^2}\right)-G\left(\sqrt{(x-x')^2+(y+y')^2}\right).
\end{eqnarray}
The Green function obeys the boundary condition $G_+(x,y|x',y'=0)=0$.
The components of the metric tensor (\ref{mt}) are easily obtained
\begin{eqnarray}
\label{metricwall}
g_{ij}(x,y)&=&\left<\partial_i\phi(x,y)\partial_j\phi(x,y)\right>:\nonumber\\
g_{xx}&=&-G''(0)+\frac{G'(2y)}{2y},\nonumber\\
g_{yy}&=&-G''(0)-G''(2y),\nonumber\\
g_{xy}=g_{yx}&=&0.
\end{eqnarray}
The signed density of extremal points is the scalar curvature of a two-dimensional manifold with metric tensor (\ref{metricwall})
\begin{equation}
4\pi\rho(y)=\sqrt{\det g}R
\end{equation}
where the scalar curvature is twice the Gaussian curvature $R=2K$ in two dimensions.
The metric is orthogonal and depends on the distance to the wall $y$ only.
The non-vanishing affine connections are
\begin{eqnarray}
\left<\partial_x\partial_y\phi\partial_x\phi\right>&=&\partial_yg_{xx}/2\nonumber\\
\left<\partial_y\partial_y\phi\partial_y\phi\right>&=&\partial_yg_{yy}/2\nonumber\\
\left<\partial_x\partial_x\phi\partial_y\phi\right>&=&-\partial_yg_{xx}/2
\end{eqnarray}
since $\left<\partial_x\phi\partial_y\phi\right>=0$ and, therefore, $\left<\partial_x\partial_x\phi\partial_y\phi\right>+\left<\partial_y\partial_x\phi\partial_x\phi\right>=0$.
The scalar curvature is according to \cite{Fol03a}
\begin{eqnarray}
\fl R&=&(\det g)^{-1}
e_{ij}e_{kl}\bigl(\left<\partial_i\partial_k\phi\partial_j\partial_l\phi\right>-
\left<\partial_i\partial_k\phi\partial_m\phi\right>g^{mn}\left<\partial_n\phi\partial_j\partial_l\phi\right>\bigr)\nonumber\\
&=&\frac{2}{g_{xx}g_{yy}}\left(-\frac{1}{2}\partial_y^2g_{xx}+\frac{1}{4}\partial_yg_{xx}\left((g_{yy})^{-1}\partial_yg_{yy}+(g_{xx})^{-1}\partial_yg_{xx}\right)\right)
\end{eqnarray} where $e_{12}=1, e_{21}=-1, e_{11}=e_{22}=0$.
A straightforward calculation yields
\begin{equation}
\label{walldens}
4\pi\rho(y)=-\partial_y\left(\left(g_{xx}g_{yy}\right)^{-1/2}\partial_yg_{xx}\right)=\partial_yf(y)
\end{equation}
where $f(y)=-\left(g_{xx}g_{yy}\right)^{-1/2}\partial_yg_{xx}$ is the integrated charge density.


\subsection{Random waves in half space}
\label{sec-halfrw}
Chaotic wave functions are well modelled by  a superposition of partial waves with Gaussian distributed amplitudes and fixed energy \cite{Ber77}. The partial waves $\phi_k$ must obey the Helmholtz equation $-\nabla^2\phi_k=k^2\phi_k$ with  $k$ being fixed. Without lost of generality we can set $k^2=1$. For time reversal symmetry, i.e. in the absence of a magnetic field, we can choose real amplitudes. The  corresponding
free-space correlation function of the $\phi$ reads \cite{Ber02,Ber77,Den01b,Ber00,Den01a,Blu02}
\begin{equation}
\label{rw-corr}
G(r)=\frac{1}{(2\pi)^2}\int\rmd^2p\,\delta(p^2-1)\exp(i\bi{p}\bi{r})\propto J_0(r),
\end{equation}
where $J_0$ is the zeroth-order Bessel function.
$G(r)$ also models the correlation of the undulations  $\phi(\bi{r})$ of capillary waves (with fixed energy) at the surface of a liquid. We might therefore also apply the following results to the extremal points of random capillary waves or even to ocean waves \cite{Lon57a,Lon57b,Lon60a,Lon60b,Lon60c}.

\begin{figure}
\includegraphics[scale=0.8]{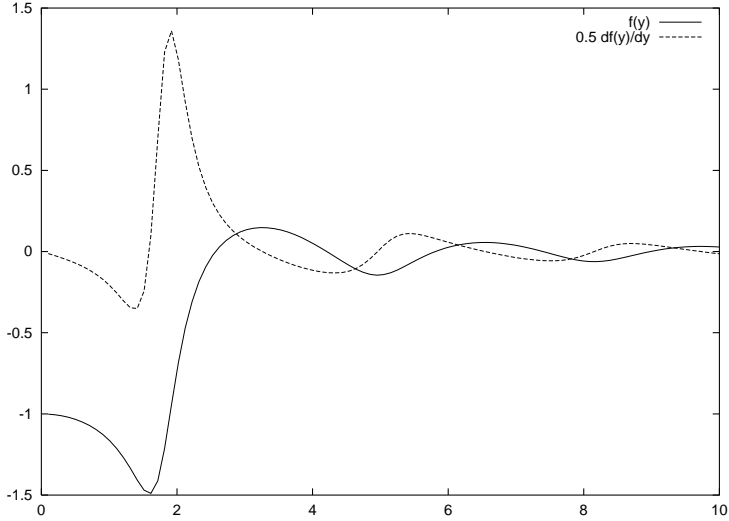}
\caption[Charge density of a chaotic wave function next to a wall]{\label{wall}Integrated charge density $f(y)$ (solid) and charge density $2\pi\rho$ (dashed) of a chaotic wave function next to a wall.}
\end{figure}

We compute the density of extrema of a random wave near a straight wall with Dirichlet boundary conditions.  We expect it to be a model for the distribution of extremal points of a chaotic wave function near the boundary \cite{Ber02}. According to section  \ref{sec-bound} we have to compute  
the nonvanishing components of the metric tensor (\ref{mt})
\begin{eqnarray}
g_{xx}&=&1/2-\frac{J_1(2y)}{2y}=\left(1-J_0(2y)-J_2(2y)\right)/2,\nonumber\\
g_{yy}&=&1/2+J_1'(2y)=\left(1+J_0(2y)-J_2(2y)\right)/2
\end{eqnarray}
which can be found in  \cite{Ber02}.
We obtain $\partial_yg_{xx}=2J_2(2y)/(2y)$ and the integrated charge density (\ref{walldens})
\begin{equation}
 \fl f(y)=\frac{-\partial_yg_{xx}}{\sqrt{g_{xx}g_{yy}}}=\frac{-4J_2(2y)}{2y\left(\left(1-J_2(2y)\right)^2-\left(J_0(2y)\right)^2\right)^{1/2}}=-(1+y^2/8+\ldots).
\end{equation}
Figure \ref{wall} shows $f(y)$ and its derivative $2\pi\rho=\partial_yf(y)/2$. Next to the wall one observes a negative density, i.e. an excess of saddle points, followed by a sharp positive peak. The system has, contrary to the free one, a net excess charge $4\pi\int\rmd y\rho=f(0)=1/2$ per unit length.

\subsection{A surface of revolution}

\begin{figure}
\includegraphics[scale=0.8]{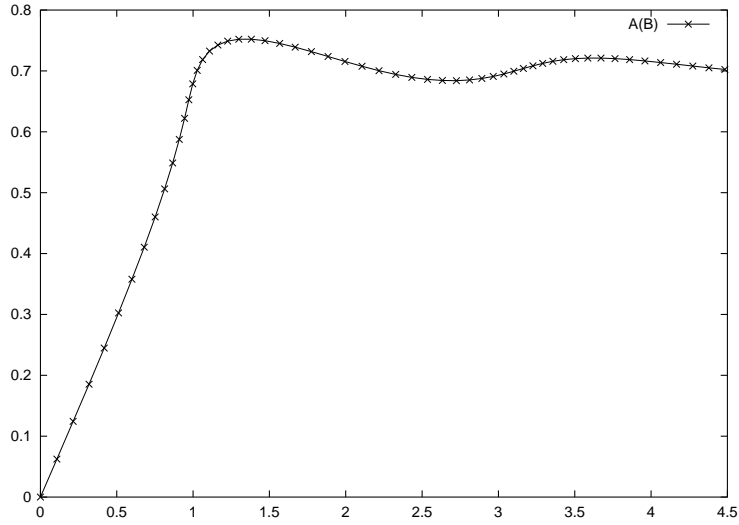}
\caption[Contour of the surface of revolution]{\label{sorcut}Contour of the surface of revolution. Shows $A(B)$ for $0\le y<7$. Two consecutive crosses correspond to an interval of $\Delta y=1/8$.}
\end{figure}

\begin{figure}
\includegraphics[scale=1]{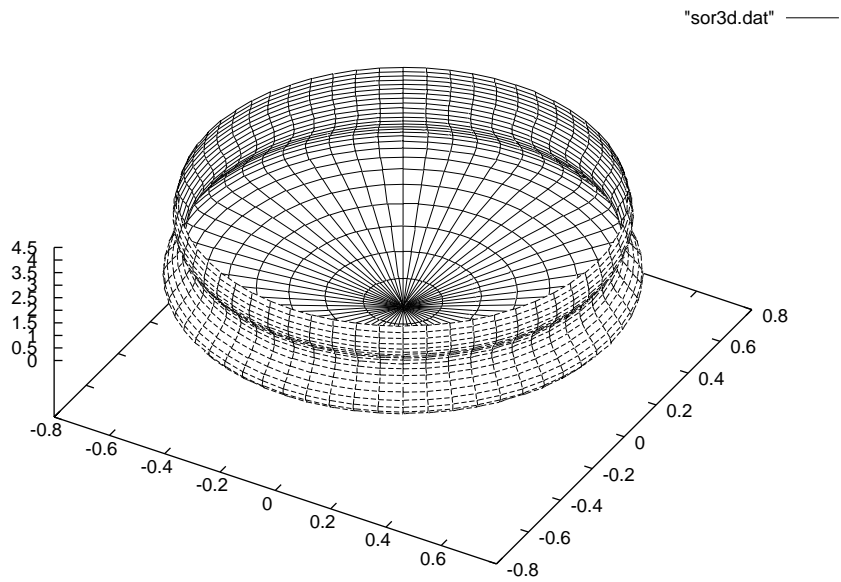}
\caption[Surface of revolution]{\label{sor3d_30}Surface of revolution. Shown for $0\le y<7$. Two consecutive meridian correspond to an interval of $\Delta y=1/4$.}
\end{figure}

\begin{figure}
\includegraphics[scale=1]{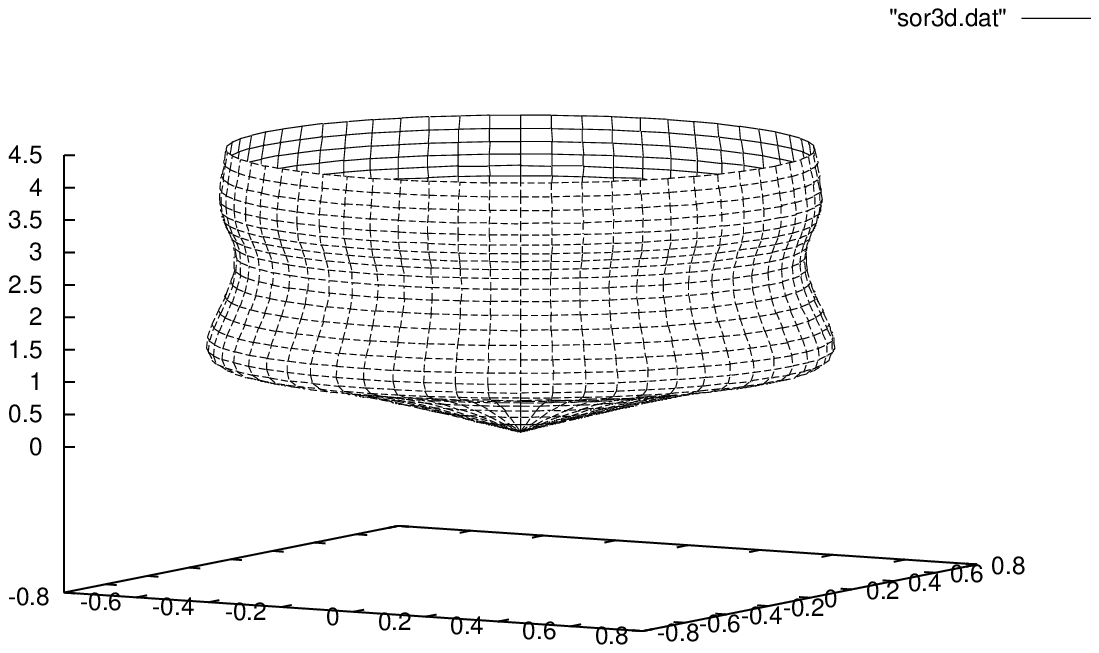}
\caption[Different view on the surface of revolution]{\label{sor3d_80}Different view on the surface of revolution. Shown for $0\le y<7$. Two consecutive meridian correspond to an interval of $\Delta y=1/4$.}
\end{figure}

\begin{figure}
\includegraphics[scale=0.8]{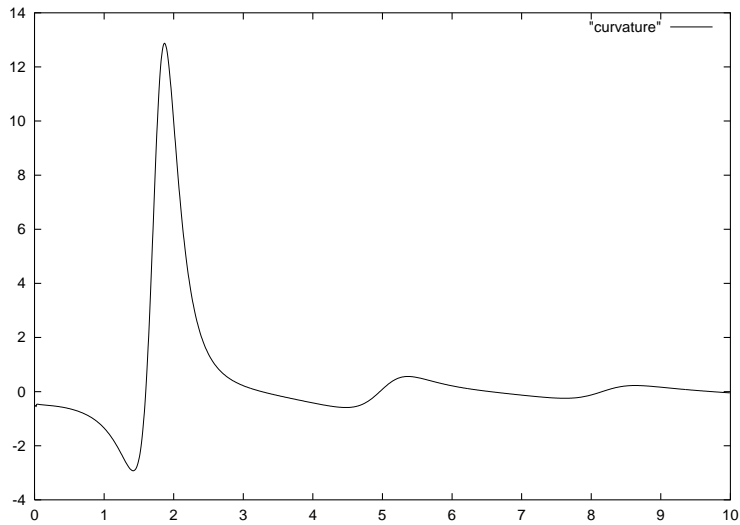}
\caption{\label{fig-curv}Scalar curvature $R(y)$ for random waves.}
\end{figure}

There is a good way to visualize the charge density (\ref{walldens}). It is possible to embed the manifold with metric tensor (\ref{metricwall}) in three-dimensional space. This allows us to literally see the defect density, which is the curvature of the embedded surface times the element of area $\sqrt{\det g}$. There is, however, a price to be paid -- we must compactify the dimension $x$ by making it $2\pi$-periodic.
Then one can construct a surface of revolution with metric tensor (\ref{metricwall}). The surface of revolution is parametrized through a coordinate $y$ along the principal axis and the angle $0<x\le 2\pi$
\begin{equation}
\label{wall-sor}
\vec{X}(x,y)=\left(\begin{array}{c}A(y)\cos x\\ A(y)\sin x\\B(y)\end{array}\right)
\end{equation}
with $A>0$ and a monotonically growing $B(y)$. $\vec{X}$ is a vector in three-dimensional, Euclidean space. For a compact  overview of the differential geometry of surfaces, especially of embedded surfaces, see e.g. \cite{Dav89,Kam02}.
The (induced) metric tensor of the surface of revolution reads
\begin{eqnarray}
g_{xx}&=&\partial_x\vec{X}\cdot\partial_x\vec{X}=A^2(y),\nonumber\\
g_{xy}&=&0,\nonumber\\
g_{yy}&=&\partial_y\vec{X}\cdot\partial_y\vec{X}=(\partial_yA)^2+(\partial_yB)^2.
\end{eqnarray}
Thus, the above metric is equal to the metric (\ref{metricwall}) provided we identify
\begin{eqnarray}
A(y)&=&\sqrt{g_{xx}},\nonumber\\
\partial_yB(y)&=&\left(g_{yy}-(\partial_yA)^2\right)^{1/2}=\sqrt{g_{yy}}\sqrt{1-f^2(y)/4}\nonumber\\
B(y)&=&\int_{0}^{y}\rmd\bar{y}\sqrt{g_{yy}(\bar{y})(1-f^2(\bar{y})/4)}.
\end{eqnarray}
It is important to keep the information about the coordinate $y$, e.g. by displaying a grid with lines at equidistant intervals in $y$-space to be able to reconstruct the element of area $\sqrt{\det g}$. 
The  embedding (\ref{wall-sor}) of the metric tensor (\ref{mt}) is restricted to regions, where the integrated charge density obeys $|f(y)|<2$.
The contour of the surface of revolution for the case of random waves is shown in figure \ref{sorcut}, where the function $A(B)$ for 
$0\le y<7$ is displayed. The crosses display the $y$-coordinate--two consecutive crosses represent an interval $\Delta y=1/8$.
Figures \ref{sor3d_30} and \ref{sor3d_80} show two projections of the three-dimensional surface. The meridians are at values of $y=k/4, k=0,1,2,\ldots, 0\le y<7$. To obtain $4\pi\rho$ for a given $y$, one has to find the meridian of number $4\times y$ (counted from the cone), the local Gaussian curvature there and multiply it with $\sqrt{\det g}$, which is proportional to the area of the corresponding grid element.
The surface of revolution has a conical form for $y\rightarrow 0$, which is an artefact of the embedding (\ref{wall-sor}). In fact, expanding the surface for small $y$ we obtain a conical geometry: $A(y)=y/2+\ldots$ and $B(y)=\sqrt{3}y/2 +\ldots$. The curvature itself is negative at the origin and has the limiting value $R=f'(y)(g_{xx}g_{yy})^{-1/2}\rightarrow -1/2$. Figure \ref{fig-curv} displays the curvature as a function of the coordinate $R=R(y)$. Note the pronounced positive peak of the scalar curvature at $y\approx 2$.


\subsection{Pinned fluctuating surfaces}
\begin{figure}
\includegraphics[scale=0.8]{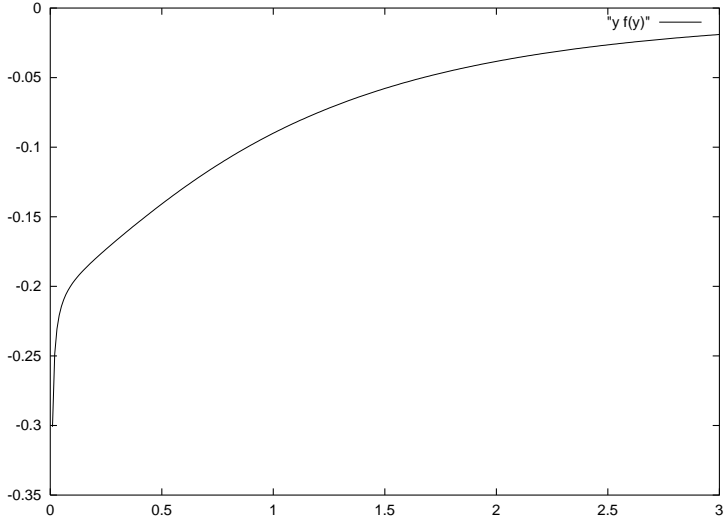}
\caption{\label{wall2}The function $y f(y)$ for membranes.}
\end{figure}

\begin{figure}
\includegraphics[scale=0.8]{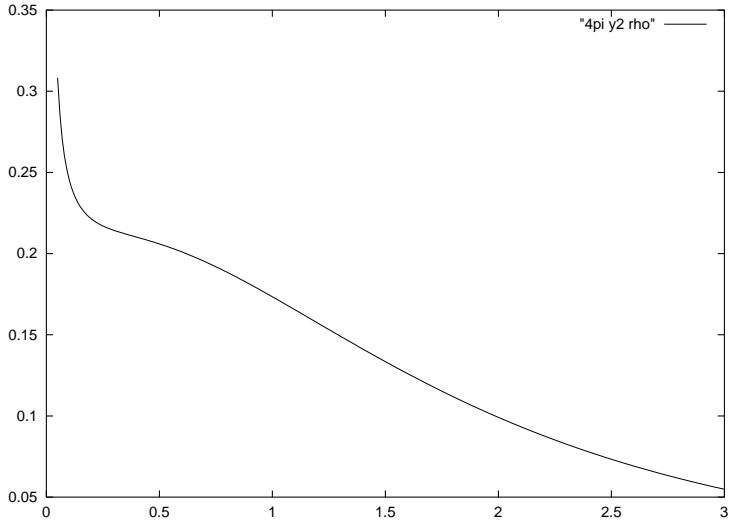}
\caption{\label{wall3}The function $4\pi y^2 \rho$ for membranes.}
\end{figure}

As a further application, we calculate the density of extrema for a nearly planar, fluctuating fluid membrane, attached to a straight line $y=0$ (imagine a membrane suspended in a large frame). We represent the shape of the membrane by a height variable $\phi(\bi{r})$. The domain of $\bi{r}$ is restricted to the upper half space $y>0$, the boundary condition is $\phi(x,y=0)=0$.
The thermal shape fluctuations of the fluid membrane are described by the Helfrich-Hamiltonian \cite{Hel73,Pel85,Nel89}
\begin{equation}
H/T=\int\rmd^2r\left(\kappa(\nabla^2\phi)^2+\sigma(\nabla\phi)^2\right)
\end{equation}
plus higher-order terms. $\kappa$ is the so-called bending rigidity, $\sigma$ is the effective surface tension.  For almost planar surfaces the fluctuations of the height variable $\phi$ are  Gaussian to lowest order. We rescale the height $\sqrt{\kappa}\phi\rightarrow\phi$ and set
$\tau=\sigma/\kappa$. The correlation function for the unconstrained $\phi$ is now
\begin{equation}
\fl G(r)=\frac{1}{(2\pi)^2}\int\rmd^2p\frac{\exp(i\bi{p}\bi{x})}{p^2(p^2+\tau)}=\frac{1}{2\pi}\left(C+ \log(1/r)-K_0(r\sqrt{\tau})\right),
\end{equation}
where $C$ is an irrelevant constant.
We rescale the length scale that $\tau=1$ and regularize $G$ at distances $r<a$.
With the help of $K_0'(r)=-K_1(r)$ and $K_1'(r)=-(1/2)(K_0(r)+K_2(r))$ and
$K_1(r)/r=-(1/2)(K_0(r)-K_2(r))$ we obtain the components of the metric tensor (\ref{metricwall})
\begin{eqnarray}
g_{xx}&=&B-(2y)^{-2}+K_2(2y)/2-K_0(2y)/2\nonumber\\
g_{yy}&=&B-(2y)^{-2}+K_2(2y)/2+K_0(2y)/2,
\end{eqnarray}
where $B=-G''(0)$. For distances $y$ much larger than the correlation length (which is $1$ in our units) we have $g_{xx}=g_{yy}\approx B-(2y)^{-2}$. The corresponding charge density (\ref{walldens}) has an algebraic decay
\begin{equation}
4\pi\rho\sim\frac{3}{2B}y^{-4}\textrm{ for }r\gg 1,
\end{equation}
a reminiscence of the long ranged interactions (from the $1/p^2$-term in the Green function).
The integrated charge density $f(y)$ is
\begin{equation}
\label{denswall}
\fl f(y)= \frac{-1/(2y^3)+K_2(2y)/y}{\sqrt{g_{xx}g_{yy}}}=
\frac{-1/(2y^3)+K_0(2y)/y+K_1(2y)/y^2}{\sqrt{g_{xx}g_{yy}}}.
\end{equation}
For small arguments $a<y\ll 1$ the integrated charge density behaves as $f(y)\sim y^{-1}\tilde{F}(\log y)  +\Or(y^{-2})$. Figure \ref{wall2} shows $y f(y)$.
Figure \ref{wall3} displays the function $4\pi y^2\rho$, where again $y^{-2}$ is the leading singular behaviour of the charge density for small arguments.
The cutoff region $0<y\lesssim a$ is not shown in both figures. 


\section{The two-point function}
\label{sec-cccg}
We present a convenient way of calculating the two-point function (charge-charge correlation function) $C(r)=\left<\rho(0)\rho(\bi{r})\right>$.
We have mentioned in the introduction that the two-point function  can be expressed through the Riemannian curvature tensor $R_{k\gamma,l\lambda,i\alpha,j\beta}$ of a four-dimensional Riemann manifold with metric tensor $\left<\partial_i\phi(\bi{r}_\alpha)\partial_j\phi(\bi{r}_\beta)\right>$, where $\alpha=1,2,\,\beta=1,2$ enumerate the two points. The Riemann curvature tensor is evaluated in \ref{sec-einstein}. The derivation of the correlation function with the help of differential geometry, however, is rather involved. Instead we utilize a less tedious approach, which is even valid for more general Gaussian vector fields $u_i, i=1,2$ with an isotropic, translationally invariant correlation function
\begin{equation}
\left<u_i(\bi{r})u_j(\bi{r}')\right>=\chi_{ij}(\bi{r}-\bi{r}').
\end{equation}
In \ref{app-cccf} we obtain for the charge-charge correlation function
\begin{equation}
(2\pi)^2\left<\rho(\bi{r})\rho(0)\right>=(2\pi)^2C(r)=-\partial_i\partial_j\Omega_{ij}
\end{equation}
where the potential $\Omega_{im}$ reads
\begin{equation}
\Omega_{im}(\bi{r})=e_{ij}e_{mn}\frac{1}{2}(\det h)^{-1/2}e_{ab}e_{cd}\partial_j\chi_{ac}\partial_n\chi_{bd}
\end{equation}
with $h_{ij}=\left<u_iu_m\right>\left<u_ju_m\right>-\chi_{im}\chi_{jm}$.
Summation over double indices is implied.
Next, we restrict the vector field $u_i$ to gradient fields $u_i=\partial_i\phi$. The nodal points of $u_i$, i.e. the points where $u_i$ vanishes, are the extremal point of the scalar field $\phi$. The potential becomes
\begin{equation}
\Omega_{im}(\bi{r})=e_{ij}e_{mn}\frac{1}{2}(\det h)^{-1/2}e_{ab}e_{cd}\partial_j\partial_a\partial_cG\partial_n\partial_b\partial_dG
\end{equation}
and $h_{ij}=\delta_{ij}-\partial_i\partial_mG\partial_m\partial_jG$. For convenience, we have chosen the amplitude of $\phi$ that $-G''(0)=1$. Therefore, it is $-\partial_i\partial_jG(0)=\delta_{ij}$.
As a consequence of the polar invariance, matrices (such as $\partial\partial G$, $\Omega$ and $h$) have the general form
\begin{equation}
\label{radmatrix}
A_{ij}=A_{11}\frac{r_ir_j}{r^2}+A_{22}\left(\delta_{ij}-\frac{r_ir_j}{r^2}\right).
\end{equation}
For the particular choice $\bi{r}=(r,0)$ the only remaining components are $A_{11}$ and $A_{22}$. 
The non-vanishing derivatives of order 2, 3, 4 of $G(r)$ and components of $h$ and $\xi$ for $\bi{r}=(r,0)$ are given in equation (\ref{partial})  (the prime $'$ denotes the derivative with respect to $r$)
\begin{eqnarray}
\label{partial}
\partial_1\partial_1G&=&Z_1=G''(r)\nonumber\\
\partial_2\partial_2G&=&Z_2=G'(r)/r\nonumber\\
\partial_1\partial_1\partial_1G&=&Z_1'\nonumber\\
\partial_1\partial_2\partial_2G&=&Z_2'=(Z_1-Z_2)/r\nonumber\\
\partial_1\partial_1\partial_1\partial_1G&=&Z_1''\nonumber\\
\partial_1\partial_1\partial_2\partial_2G&=&Z_2''\\
\partial_2\partial_2\partial_2\partial_2G&=&3Z_2'/r\nonumber\\
h_{11}&=&D_1=1-Z_1^2\nonumber\\
h_{22}&=&D_2=1-Z_2^2\nonumber\\
\xi_{11}&=&Z_1/D_1\nonumber\\
\xi_{22}&=&Z_2/D_2\nonumber\end{eqnarray}
A short calculation yields
\begin{equation}
\Omega_{11}=-(Z_2')^2(D_1D_2)^{-1/2},\,\,
\Omega_{22}=Z_1'Z_2'(D_1D_2)^{-1/2}
\end{equation}where
$Z_1=G''(r), Z_2=G'/r, D_1=1-Z_1^2, D_2=1-Z_2^2$.
We find for a matrix of the form (\ref{radmatrix})
\begin{equation}
\partial_i\partial_jA_{ij}=\nabla^2A_{11}+\frac{1}{r}\partial_r\left(A_{11}-A_{22}\right),
\end{equation} where $\nabla^2A_{11}=(rA_{11}')'/r$.
The correlation function becomes
\begin{equation}
\label{tp-func}
(2\pi)^2C(r)=-\partial_i\partial_j\Omega_{ij}=-\nabla^2\Omega_{11}-\left(\Omega_{11}-\Omega_{22}\right)'/r\equiv\psi'/r,
\end{equation} with the `potential'
\begin{eqnarray}
\label{tp-pot}
\psi(r)&=&\Omega_{22}-\Omega_{11}-r\Omega_{11}'=\Omega_{22}-(r\Omega_{11})'\nonumber\\
&=&\frac{3(Z_1'-Z_2')(Z_1-Z_2)}{r(D_1D_2)^{1/2}}+\frac{(Z_1-Z_2)^2}{r}\left((D_1D_2)^{-1/2}\right)'\nonumber\\
&=&\frac{1}{r(Z_1-Z_2)}\left(\frac{(Z_1-Z_2)^3}{(D_1D_2)^{1/2}}\right)'.
\end{eqnarray}
This result was also obtained by \cite{Den03a} using a direct approach. 
The correlation function of  the Gaussian field $\phi$ can now be written as 
\begin{equation}
G(r)=\int\rmd p_1\rmd p_2 G\left(\sqrt{p_1^2+p_2^2}\right)\cos(p_1 r)\equiv\int\rmd p\tilde{G}(p)\cos(pr)
\end{equation} with $G(p)>0$ and also $\tilde{G}(p)>0$. It is
\begin{equation}
1=-G''(r=0)=\int\rmd p\tilde{G}(p)p^2,
\end{equation}
i.e. $w(p)=p^2\tilde{G}(p)$ has the properties of a probability density. The $r$-expansion of $G(r)$ reads
\begin{equation}
G(r)= -\frac{r^2}{2} + \frac{b\,r^4}{4!} - \frac{c\,r^6}{6!} + \frac{d\,r^8}{8!} - 
   \frac{e\,r^{10}}{10!}+\Or(r^{12}),
\end{equation}
where $b=G''''(r=0)=\int\rmd p\,w(p)p^2$ and $c=-G^{(6)}(r=0)=\int\rmd p\,w(p)p^4$ (as well as $d$ and $e$) are positive constants.
With the help of Mathematica$^\textrm{\tiny TM}$ we find the following series expansion of $\psi(r)$:
\begin{eqnarray}
\label{psi-exp}
\psi(r)&=&\frac{4\,{ b }}{3\,{\sqrt{3}}} + 
   \frac{\left( b^2 - c \right) \,r^2}{3\,{\sqrt{3}}}\nonumber\\
&&{}+ 
   \frac{\left( 45\,b^4 - 56\,b^2\,c + 3\,c^2 + 10\,b\,d \right) \,r^4}
    {540\,{\sqrt{3}}\,{ b }} + \Or(r^6).
\end{eqnarray}
It is shown in equation (\ref{grad-sumr}) that $-\psi(0)$ is related to the absolute density of extrema $n_0=\left<|\rho|\right>$. The next term of the $\psi$ expansion determines the charge-charge correlation function at almost coinciding points\footnote{$C(0)$ is not well defined, since $C(r)$ has a $\delta$-like singularity at the origin, see \cite{Fol03a}. The $\delta$-function, however, is lost `somewhere' during the course of section \ref{sec-cccg}. A careful analysis at the origin similar to the one presented in \cite{Fol03a} is needed to recover the $\delta$-function.}
\begin{eqnarray}
(2\pi)^2\lim_{r\rightarrow 0}C(r)&=&\frac{-2(c-b^2)}{3\sqrt{3}}\nonumber\\
&=&\frac{-2}{3\sqrt{3}}\left(\int\rmd p\,w(p)p^4-\left(\int\rmd p\,w(p)p^2\right)^2\right).
\end{eqnarray}
The latter expression is obviously negative--the vicinity of an extremum is populated by extrema of the opposite type \cite{Shv94b,Fre94b}, which screen the central charge. 

\subsection{Perfect screening of gradient fields}
\label{sec-grsr}
We show that the charge-charge correlation function $C(r)$ obeys the first Stillinger-Lovett sum rule
\begin{equation}
\lim_{\epsilon\rightarrow 0}\int_{|\bi{r}|>\epsilon}\rmd^2rC(r)=-n_0.
\end{equation}
It expresses the fact that a particular extremum is \textit{completely} screened by extrema (charges) of the opposite sign \cite{Ber00,Sti68a,Sti68b,Hal81,Mar83,Liu92,Fre98}. 
Applying the results of section \ref{sec-cccg} we find for the left hand side
\begin{equation}
\label{grad-sumr}
\fl -n_0=2\pi\int_{0^+}^\infty r\rmd r C(r)=\frac{1}{2\pi} \int_{0^+}^\infty\rmd r\psi'(r)=-\frac{1}{2\pi}\psi(0)=-\frac{2G''''(r=0)}{3\pi\sqrt{3}}
\end{equation}
where $G(r)$ is the correlation function of the field $\phi$. On the other hand, the absolute density of extremal points of the Gaussian field $\phi$ reads \cite{Lon57a,Lon57b} (for an arbitrary point $\bi{r}$)
\begin{equation}
n_0=\left<|\partial_x\partial_x\phi\partial_y\partial_y\phi-(\partial_x\partial_y\phi)^2|\delta(\partial_x\phi)\delta(\partial_y\phi)\right>.
\end{equation}
The Gaussian variables $\partial_i\phi$ are independent of the second derivatives $\partial_i\partial_j\phi$ (evaluated at the same point). It is therefore $n_0=\left<|\ldots|\right>\left<\delta(\partial_x\phi)\delta(\partial_y\phi)\right>$ with 
\begin{eqnarray}
\left< \delta(\partial_x\phi)\delta(\partial_y\phi)\right>&=&\int\rmd^2E\frac{1}{\pi\left<(\nabla\phi)^2\right>}\exp\left(\frac{E^2}{\left<(\nabla\phi)^2\right>}\right)\delta(E_x)\delta(E_y)\nonumber\\
&=&\frac{1}{\pi\left<(\nabla\phi)^2\right>}=1/(2\pi),
\end{eqnarray}
where we have chosen $-G''(0)=1$.
The second moments of $\partial_i\partial_j\phi$ at coinciding points are
\begin{eqnarray}
\left<\partial_x\partial_y\phi\partial_x\partial_x\phi\right>&=&
\left<\partial_x\partial_y\phi\partial_y\partial_y\phi\right>=0\nonumber\\
\left<\partial_x\partial_x\phi\partial_x\partial_x\phi\right>&=&\left<\partial_y\partial_y\phi\partial_y\partial_y\phi\right>=G''''(0)\nonumber\\
\left<\partial_x\partial_y\phi\partial_x\partial_y\phi\right>&=&\left<\partial_x\partial_x\phi\partial_y\partial_y\phi\right>=G''''(0)/3.
\end{eqnarray}
Next, we define the mutually independent Gaussian variables
$X=(\partial_x\partial_x\phi+\partial_y\partial_y\phi)/\sqrt{2}$, $Y=(\partial_x\partial_x\phi-\partial_y\partial_y\phi)/\sqrt{2}$, and $Z=\partial_x\partial_y\phi$ with
$\left<X^2\right>=(4/3)G''''(0)$, $\left<Y^2\right>=(2/3)G''''(0)$, and $\left<Z^2\right>=(1/3)G''''(0)$.
Then
\begin{eqnarray}
\left<|\ldots|\right>&=&\left<\left|X^2/2-Y^2/2-Z^2\right|\right>\nonumber\\
&=&\frac{1}{(2\pi)^{3/2}}\int\rmd x\rmd y\rmd z\exp\left(-(x^2+y^2+z^2)/2\right)\nonumber\\
&&\times\left|\frac{2G''''(0)}{3}x^2-\frac{G''''(0)}{3}(y^2+z^2)\right|.
\end{eqnarray}
We introduce spherical coordinates $y=r\sin\theta\cos\varphi$, $z=r\sin\theta\sin\varphi$, and $x=r\cos\theta$, yielding
\begin{eqnarray}
\fl \left<|\ldots|\right>&=&G''''(0)\,\frac{2}{3\sqrt{2\pi}}\int_0^\infty\rmd r\,r^4\exp(-r^2/2)\,\int_0^{\pi/2}\rmd\theta\,\sin\theta|3(\cos\theta)^2-1|\nonumber\\
&=&\frac{4G''''(0)}{3\sqrt{3}}.
\end{eqnarray}
Finally, we obtain the absolute density of extremal points of $\phi$,
\begin{equation}
n_0=\frac{2G''''(0)}{3\pi\sqrt{3}} 
\end{equation}
in agreement with equation (\ref{grad-sumr}).

\subsection{Generating functionals for the two-point function} 

We have found a representation of the two-point function $C(r)$ as a functional derivative. 
Although this section is less useful from the practical point of view, it highlights an important mathematical aspect of the theory and demonstrates once more the applicability  of the differential geometrical approach. 
We begin with the integral over the two-point function times the correlation function of the field $\phi$,
\begin{eqnarray}
\label{intcg}
\lefteqn{-2\pi\int\rmd^2rC(r)G(r)=(2\pi)^{-1}\int\rmd^2r \partial_i\partial_j\Omega_{ij}G}\nonumber\\
&=&\textrm{b.c.}+(2\pi)^{-1}\int\rmd^2r \Omega_{ij}\partial_i\partial_jG\nonumber\\
&=&\textrm{b.c.}+\int r\rmd r(\Omega_{11}Z_1+\Omega_{22}Z_2)\nonumber\\
&=&\textrm{b.c.}+\int r\rmd r\frac{1}{(D_1D_2)^{1/2}}(Z_1'Z_2'Z_2-Z_2'Z_2'Z_1)\nonumber\\
&=&\textrm{b.c.}+\int\rmd r\frac{Z_1-Z_2}{(D_1D_2)^{1/2}}\left(Z_1'Z_2-Z_2'Z_1\right)\nonumber\\
&=&-\int\rmd r\frac{D}{(D_1D_2)^{1/2}}(Z_1'+Z_2')\nonumber\\
&&{}+\int\rmd r\left((D_2/D_1)^{1/2}Z_1'+(D_1/D_2)^{1/2}Z_2'\right)\nonumber\\
&=&\mathcal{H}-\mathcal{L}
\end{eqnarray}
where b.c. represents contributions from the boundaries $r=0, r\rightarrow\infty$, and 
\begin{eqnarray}
\mathcal{H}&=&-\int\rmd r\frac{D}{(D_1D_2)^{1/2}}(Z_1'+Z_2'),\nonumber\\
\mathcal{L}&=&-\int\rmd r\left((D_2/D_1)^{1/2}Z_1'+(D_1/D_2)^{1/2}Z_2'\right)
\end{eqnarray} with $D=1-Z_1Z_2$.
The action $\mathcal{H}$ is identified in \ref{sec-einstein} as the
integral over the scalar curvature\footnote{In principle, one has to integrate over four coordinates. The metric, however, depends on the distance $|\bi{r}^A-\bi{r}^B|$ only. The integration over the centre of mass is therefore trivial.} $R=g^{k\gamma,i\alpha}g^{l\lambda,j\beta}R_{k\gamma,l\lambda,i\alpha,j\beta}$
\begin{equation}
\label{hilbert}
\mathcal{H}=(4\pi)^{-1}\int\rmd^2r\sqrt{\det g}R.
\end{equation}
This functional is well known from general relativity. The curved spacetime manifolds which obey the least-action principle $\delta\mathcal{H}=0$ are the solutions of the Einstein field
equations (without matter).
We show below that action (\ref{hilbert}) is also  meaningful for our manifold.

The action $\mathcal{L}$ is the generating functional of $C(r)$ as computed in \ref{sec-lagrangian}
\begin{equation}
\frac{1}{r}\frac{\delta\mathcal{L}}{\delta G}=\psi'/r=(2\pi)^2C(r).
\end{equation}
The Legendre transform of $\mathcal{L}$ reads, using (\ref{intcg}) and the above equation
\begin{equation}
\mathcal{L}-\int\rmd^2r\frac{\delta\mathcal{L}}{\delta G(\bi{r})}G(\bi{r})=\mathcal{L}-\int\rmd r\frac{\delta\mathcal{L}}{\delta G(r)}G(r)=\mathcal{H}.
\end{equation}
The `Einstein'-action $\mathcal{H}$, expressed as a functional of the charge-charge correlation function $C(r)$ is therefore the generating functional of the conjugated field-field correlation function $G(r)$,
\begin{equation}
\frac{1}{(2\pi)^2r}\frac{\delta\mathcal{H}[C]}{\delta C(r)}=-G(r).
\end{equation}

\subsection{Two point function for random waves}
\label{sec-chaos}
\begin{figure}
\includegraphics[scale=1]{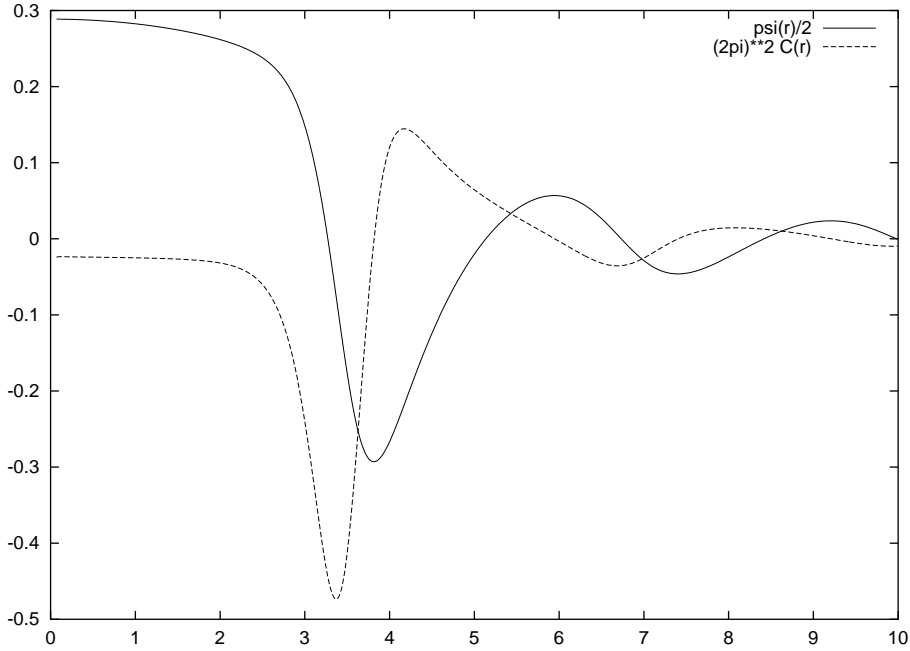}
\caption[The charge-charge correlation function]{\label{fig-chaos2p}The charge-charge correlation function $(2\pi)^2C(r)$ (dashed) and its potential $\psi(r)$ (solid; $\psi/2$ is shown for scaling reasons).}
\end{figure}

We now compute as an application the two-point function $C(r)$ for Gaussian random waves $\phi$ (without boundary).
We rescale the correlation function  (\ref{rw-corr}) of the field $\phi$ for practical reasons and use $G(r)=\left<\phi(\bi{r})\phi(0)\right>=2J_0(r)$. It is $-G''(0)=1$, $Z_1=G''(r)=-J_0(r)+J_2(r)$, $Z_2=-(J_0(r)+J_2(r))$, $D_1=1-(J_0(r)-J_2(r))^2$, $D_2=1-(J_0(r)+J_2(r))^2$, yielding for the potential (\ref{tp-pot})
\begin{equation}
\psi(r)=\frac{4}{rJ_2(r)}\frac{\rmd}{\rmd r}\left(\frac{(J_2)^3}{\left(1-(J_0-J_2)^2\right)^{1/2}\left(1-(J_0+J_2)^2\right)^{1/2}}\right).
\end{equation}
With the help of the relations $J_0'=-J_1$, $J_2'=J_1-2J_2/r$ and $2J_1/r=J_0+J_2$ we obtain the result 
\begin{eqnarray}
\psi(r)&=&\frac{2(J_2)^3}{\left(1-(J_0-J_2)^2\right)^{1/2}\left(1-(J_0+J_2)^2\right)^{1/2}}
\nonumber\\
&&\times\left(\left(1+\frac{J_0}{J_2}\right)\left(\frac{3}{J_2}+\frac{2(J_2-J_0)}{1-(J_2-J_0)^2}\right)\right.\nonumber\\
&&\left.{}-\frac{4}{r^2}\left(\frac{3}{J_2}+\frac{J_2-J_0}{1-(J_2-J_0)^2}+\frac{J_2+J_0}{1-(J_2+J_0)^2}\right)\right).
\end{eqnarray}
The small argument expansion of $\psi$ reads
\begin{equation}
\psi(r)=\frac{1}{\sqrt{3}}\left(1-\frac{r^2}{48}-\frac{r^4}{2304}-\frac{139 r^6}{29859840}\right)+\Or(r^8).
\end{equation}
The small $r$ behaviour of the charge-charge correlation function (\ref{tp-func}) is therefore
\begin{equation}
(2\pi)^2C(r)=-\frac{1}{24\sqrt{3}}\left(1+\frac{r^2}{24}+\frac{139r^4}{207360}\right)+\Or(r ^6).
\end{equation}
Figure \ref{fig-chaos2p} shows $\psi(r)/2$ and $(2\pi)^2C(r)$ for $0<r\le 10$. 
The plateau of the correlation function for small radii is almost perfect due to the smallness of the higher order expansion coefficients. Note  the pronounced, negative peak at $r\approx 3.4$.

\section{Conclusion}
We have studied the signed density of extremal points of scalar field with a Gaussian distribution. We have calculated the average signed density in half space with Dirichlet boundary conditions. The density is expressed as the Gaussian curvature of an associated (abstract) manifold, depending on the correlations of the scalar field. The abstract manifold could be visualized by embedding it into three-dimensional space. Then we computed the two-point correlation function of the above density, which is in principle the total curvature of a four-dimensional manifold. We identified the `Einstein'-action as the generating functional, which  relates the correlation function of the scalar field and the correlation function of the density of extrema.   Currently, no application of this equation is known. Nevertheless, we hope that its further study will ultimately lead to a deeper understanding of the differential geometry of the abstract manifold, which governs the distribution of the extremal points.    

\ack

I would like to thank H K Janssen and U Smilansky for encouraging discussions.
This work has been supported by the Deutsche Forschungsgemeinschaft
under SFB 237.


\appendix
\section{Computation of the general charge-charge correlation function}
\label{app-cccf}
We calculate the charge-charge correlation function 
for  Gaussian vector fields $u_i, i=1,2$ with an isotropic, translationally invariant correlation function
\begin{equation}
\label{tp_corr}
\left<u_i(\bi{r})u_j(\bi{r}')\right>=\chi_{ij}(\bi{r}-\bi{r}').
\end{equation}
The charge density $\rho$ is represented as
\begin{eqnarray}
\rho&=&\frac{1}{2\pi}\partial_i\left(\frac{e_{ij}e_{kl}u_k\partial_ju_l}{u^2+\epsilon}\right)
=\frac{1}{2}e_{ij}e_{kl}\partial_iu_k\partial_ju_l\frac{\epsilon}{\pi(u^2+\epsilon)^2}\nonumber\\
&\rightarrow&\det(\partial_iu_j)\delta^2(u)\textrm{ for }\epsilon\rightarrow 0
\end{eqnarray} where $e_{12}=1, e_{21}=-1, e_{11}=e_{22}=0$.
The charge-charge correlation function $C(r)\equiv\left<\rho(\bi{r})\rho(\bi{r}')\right>$ is
\begin{eqnarray}
\label{tp_charge}
4\pi^2C(r)&=&
\frac{\partial}{\partial r_i}\frac{\partial}{\partial r'_m}\left<
e_{ij}e_{kl}e_{mn}e_{st}\frac{u_k(\bi{r})\partial_ju_l(\bi{r})u_s(\bi{r}')\partial_nu_t(\bi{r}')}{(u^2(\bi{r})+\epsilon)(u^2(\bi{r}')+\epsilon)}\right>\nonumber\\
&\equiv&{}-\partial_i\partial_m\Omega_{im}(\bi{r}-\bi{r}')
\end{eqnarray}
utilizing the translational invariance of the correlation function (\ref{tp_corr}).
With the help of the Fourier-transformation 
\begin{equation}
(u^2+\epsilon)^{-1}=\frac{1}{2\pi}\int\rmd^2p\exp(i\underline{p}\underline{u})K_0(p\sqrt{\epsilon})
\end{equation}
($K_\nu$ is the second modified Bessel function of order $\nu$) we obtain
\begin{eqnarray}
\Omega_{im}(\bi{r})&=&-
\frac{1}{4\pi^2}\int\rmd^2p\rmd^2qK_0(\sqrt{\epsilon}p)K_0(\sqrt{\epsilon}q)e_{ij}e_{kl}e_{mn}e_{st}\nonumber\\
&&\times\left.\frac{\partial}{\partial p_k}\frac{\partial}{\partial q_s}\frac{\partial}{\partial A_{jl}}\frac{\partial}{\partial B_{nt}}\right|_{A=B=0}\nonumber\\
&&\times\left<\exp\left(
A_{ij}\partial_iu_j(0)+B_{ij}\partial_iu_j(\bi{r})+i\underline{p}\underline{u}(0)+
i\underline{q}\underline{u}(\bi{r})\right)\right>\nonumber\\
&=&-
\frac{1}{4\pi^2}\int\rmd^2p\rmd^2qe_{ij}e_{kl}e_{mn}e_{st}\frac{p_k}{p^2}\frac{q_s}{q^2}\left.\frac{\partial}{\partial A_{jl}}\frac{\partial}{\partial B_{nt}}\right|_{A=B=0}\nonumber\\
&&\times\left<\exp\left(
A_{ij}\partial_iu_j(0)+B_{ij}\partial_iu_j(\bi{r})+i\underline{p}\underline{u}(0)+
i\underline{q}\underline{u}(\bi{r})\right)\right>
\end{eqnarray}
in the limit $\epsilon\rightarrow 0$. The latter expression was obtained with the help of two partial integrations and $\partial_iK_0(\sqrt{\epsilon}p)=-\sqrt{\epsilon}K_1(\sqrt{\epsilon}p)p_i/p\rightarrow -p_i/p^2$.
The Gaussian average can now be done, yielding
\begin{eqnarray}
\Omega_{im}(\bi{r})&=&-
\frac{1}{4\pi^2}\int\rmd^2p\rmd^2qe_{ij}e_{kl}e_{mn}e_{st}\frac{p_k}{p^2}\frac{q_s}{q^2}\left.\frac{\partial}{\partial A_{jl}}\frac{\partial}{\partial B_{nt}}\right|_{A=B=0}\nonumber\\
&&\times\exp\left(
-(p_ip_j+q_iq_j)\left<u_iu_j\right>/2-p_iq_j\left<u_i(0)u_j(\bi{r})\right>\right)\nonumber\\
&&\times\exp\left(
A_{ij}B_{kl}\left<\partial_iu_j(0)\partial_ku_l(\bi{r})\right>+
ip_iB_{kl}\left<u_i(0)\partial_ku_l(\bi{r})\right>\right)\nonumber\\
&&\times\exp\left(iA_{ij}q_k\left<\partial_iu_j(0)
u_k(\bi{r})\right>+\textrm{ Terms }\propto A^2,B^2\right)\nonumber\\
&=&-
\frac{1}{4\pi^2}\int\rmd^2p\rmd^2qe_{ij}e_{kl}e_{mn}e_{st}\frac{p_k}{p^2}\frac{q_s}{q^2}\nonumber\\
&&\times\left(\left<\partial_ju_l(0)\partial_nu_t(\bi{r})\right>-p_v
\left<u_v(0)\partial_nu_t(\bi{r})\right>q_w\left<\partial_ju_l(0)u_w(\bi{r})\right>\right)\nonumber\\
&&\times\exp\left(-(p_ip_j+q_iq_j)\left<u_iu_j\right>/2-p_iq_j\left<u_i(0)u_j(\bi{r})\right>\right)\nonumber\\
&=&-
\frac{1}{4\pi^2}\int\rmd^2p\rmd^2qe_{ij}e_{kl}e_{mn}e_{st}\frac{p_k}{p^2}\frac{q_s}{q^2}\nonumber\\
&&\times\left(-\partial_j\partial_n\chi_{lt}(\bi{r})+p_v
\partial_n\chi_{vt}(\bi{r})q_w\partial_j\chi_{lw}(\bi{r})\right)\nonumber\\
&&\times\exp\left(-(p_ip_j+q_iq_j)\left<u_iu_j\right>/2-p_iq_j\chi_{ij}(\bi{r})\right),
\end{eqnarray}
where $\left<u_iu_j\right>=\left<u_i(0)u_j(0)\right>=\left<u_i(\bi{r})u_j(\bi{r})\right>$.
We eliminate the second derivate $\partial_j\partial_n\chi_{lt}$ by a partial `integration'
\begin{eqnarray}
\label{tp_rotation}
\fl\Omega_{im}(\bi{r})&=&e_{mn}\partial_n\left(\frac{1}{4\pi^2}\int\rmd^2p\rmd^2qe_{ij}e_{kl}e_{st}\frac{p_k}{p^2}\frac{q_s}{q^2}\partial_j\chi_{lt}(\bi{r})\exp(\ldots)\right)\\
&&{}+\frac{1}{4\pi^2}\int\rmd^2p\rmd^2qe_{ij}e_{kl}e_{mn}e_{st}\frac{p_kp_v}{p^2}\frac{q_sq_w}{q^2}\left(\partial_j\chi_{lt}\partial_n\chi_{vw}-\partial_n\chi_{vt}\partial_j\chi_{lw}
\right)\nonumber\\
&&\times\exp\left(-(p_ip_j+q_iq_j)\left<u_iu_j\right>/2-p_iq_j\chi_{ij}\right)
\end{eqnarray}
The term (\ref{tp_rotation}) does not yield a contribution to the charge-charge correlation function $\propto -\partial_i\partial_m\Omega_{im}$ since it is a pure curl.
The term $(\partial_j\chi_{lt}\partial_n\chi_{vw}-\partial_j\chi_{lw}\partial_n\chi_{vt})$ is antisymmetric with respect to the indices $t,w$:
\begin{equation}
\partial_j\chi_{lt}\partial_n\chi_{vw}-\partial_j\chi_{lw}\partial_n\chi_{vt}=e_{tw}e_{ab}\partial_j\chi_{la}\partial_n\chi_{vb}.
\end{equation}
The $q^2$-denominator is cancelled, since $e_{tw}e_{st}q_sq_w=-q^2$, yielding
\begin{eqnarray}
\Omega_{im}(\bi{r})&\equiv&-\frac{1}{4\pi^2}\int\rmd^2p\rmd^2qe_{ij}e_{kl}e_{mn}
\frac{p_kp_v}{p^2}e_{ab}\partial_j\chi_{la}\partial_n\chi_{vb}\nonumber\\
&&\times\exp\left(-(p_ip_j+q_iq_j)\left<u_iu_j\right>/2-p_iq_j\chi_{ij}\right),
\end{eqnarray}
where we have omitted the curl (\ref{tp_rotation}). Now $\Omega_{im}$ is no longer symmetric with respect to $i,m$; nevertheless only the symmetric part of $\Omega_{im}$ contributes to the charge-charge correlation function. Equivalently, we might replace $e_{ab}\partial_j\chi_{la}\partial_n\chi_{vb}$ by the symmetrized version $e_{ab}(\partial_j\chi_{la}\partial_n\chi_{vb}+\partial_n\chi_{la}\partial_j\chi_{vb})/2$. The latter expression is antisymmetric in $l,v$ and can therefore be written as $e_{lv}e_{ab}e_{cd}\partial_j\chi_{ca}\partial_n\chi_{db}/2$. Up to an antisymmetric component, $\Omega_{im}$ reads
\begin{eqnarray}
\Omega_{im}(\bi{r})&\equiv&e_{ij}e_{mn}\frac{1}{8\pi^2}\int\rmd^2p\rmd^2q
e_{ab}e_{cd}\partial_j\chi_{ca}\partial_n\chi_{db}\nonumber\\
&&\times\exp\left(-(p_ip_j+q_iq_j)\left<u_iu_j\right>/2-p_iq_j\chi_{ij}\right)
\end{eqnarray}
Next we perform the $p,q$-integral, which is Gaussian now, and obtain finally
\begin{equation}
\label{tp_potential}
\Omega_{im}(\bi{r})=e_{ij}e_{mn}\frac{1}{2}(\det h)^{-1/2}e_{ab}e_{cd}\partial_j\chi_{ac}\partial_n\chi_{bd},
\end{equation}
where $h_{ij}=\left<u_iu_m\right>\left<u_ju_m\right>-\chi_{im}\chi_{jm}$.
As an examination, we use equations (\ref{tp_charge}) and (\ref{tp_potential}) to rederive the 
well known two-point function for the special case of vector fields with independent components $\chi_{ij}(\bi{r})=\delta_{ij}G(r)$ in two dimensions (see \cite{Fol03a,Ber00,Hal81,Liu92}). Then
\begin{equation}
\label{alt-vec}
\fl\Omega_{im}=e_{ij}e_{mn}\frac{1}{G(0)^2-(G(r))^2}\partial_jG\partial_nG=e_{ij}e_{mn}\partial_jK(r)\partial_nK(r)
\end{equation}
where $K(r)=\arcsin(G(r)/G(0))$. Plugged into (\ref{tp_corr})
we get
\begin{equation}
\fl 4\pi^2C(r)=-e_{ij}e_{mn}\partial_j\partial_mK\partial_i\partial_nK=2\det(\partial_i\partial_jK)=2K''(r)K'(r)/r
\end{equation}
in agreement with the known result.


\section{The `Einstein'-action}
\label{sec-einstein}
We will now calculate the scalar curvature $R$ and the `Einstein'-action 
$
\mathcal{H}=(4\pi)^{-1}\int\rmd^2r\sqrt{\det g}R
$
for a four-dimensional Riemannian manifold with metric $g_{i\alpha,j\beta}=-\partial_i\partial_jG(|\bi{r}^\alpha-\bi{r}^\beta|)$. In principle, one has to integrate over 4 coordinates. The metric, however, depends on the distance $|\bi{r}^A-\bi{r}^B|$ only. The integration over the centre of mass is therefore trivial.
We represent the correlation function of the field $\phi$ through its Fourier transform
\begin{equation}
\left<\phi(\bi{r}^A)\phi(\bi{r}^B)\right>=G(|\bi{r}^A-\bi{r}^B|)=\int\rmd^2p \tilde{G}(p) \exp\left(i\bi{p}
\cdot(\bi{r}^A-\bi{r}^B)\right)
\end{equation}
The $4\times 4$ metric tensor reads $(i,j\in\{1,2\}; \alpha,\beta\in\{A,B\})$
\begin{equation}
g_{i\alpha,j\beta}=\left<\partial_i\phi(\bi{r}^\alpha)\partial_j\phi(\bi{r}^\beta)\right>=
\int\rmd^2p\frac{\partial\psi^*}{\partial r^{i\alpha}}\frac{\partial\psi}{\partial r^{j\beta}},
\end{equation}
where
\begin{equation}
\psi(\bi{r}^A,\bi{r}^B;\bi{p})=\sqrt{\tilde{G}(p)}\sum_{\alpha=A,B}\exp(i\bi{p}\cdot\bi{r}^\alpha)
\end{equation}
is an immersion of the four-dimensional manifold in function space.
Explicitly, we have
\begin{equation}
g_{i\alpha,j\beta}=\delta_{\alpha\beta}\delta_{ij}-c_{\alpha\beta}\partial_i\partial_jG(r)
\end{equation}
where $c_{\alpha\beta}=1-\delta_{\alpha\beta}$. 
The inverse metric tensor is easily calculated with the help of $c_{\alpha\beta}c_{\beta\gamma}=\delta_{\alpha\gamma}$:
\begin{equation}
g^{i\alpha,j\beta}=\left(\delta_{\alpha\beta}h^{ij}+c_{\alpha\beta}\xi_{ij}\right)
\end{equation}
where $h_{ij}=\delta_{ij}-\partial_i\partial_mG\partial_m\partial_jG$, $h^{ij}=h^{-1}{}_{ij}$  and $\xi_{ik}=\partial_i\partial_jG\,h^{jk}$.
The determinant of the metric tensor is $\det g=\det h$.
The components of the affine connection are
\begin{equation}
\Gamma_{k\gamma;\,i\alpha,j\beta}=\int\rmd^2p\partial_{i\alpha}\partial_{j\beta}\psi^*\partial_{k\gamma}\psi=\delta_{\alpha\beta}e_{\alpha\gamma}\partial_i\partial_j\partial_kG(r),
\end{equation}
where the index $\alpha$ is not summed. 
The components of the Riemannian curvature tensor are calculated from
\begin{eqnarray}
\fl R_{k\gamma,l\lambda,i\alpha,j\beta}&=&\int\rmd^2p\left(\partial_{i\alpha}\partial_{k\gamma}\psi^*\partial_{j\beta}\partial_{l\lambda}\psi
\right) \nonumber\\
&&{} -\int\rmd^2p\left(\partial_{i\alpha}\partial_{k\gamma}\psi^*\partial_{m\mu}\psi\right)g^{m\mu,n\nu}\int\rmd^2p'\left(\partial_{n\nu}\psi^*
\partial_{j\beta}\partial_{l\lambda}\psi\right)\nonumber\\
&&{}-(i\alpha)\leftrightarrow(j\beta).
\end{eqnarray}
Explicitly we have 
\begin{equation}
\fl\int\rmd^2p\left(\partial_{i\alpha}\partial_{k\gamma}\psi^*\partial_{j\beta}\partial_{l\lambda}\psi\right)=\delta_{\alpha\gamma}\delta_{\beta\lambda}c_{\alpha\beta}
\partial_i\partial_j\partial_k\partial_lG(r)+\delta_{\alpha\beta\gamma\delta}Z_{ijkl}
\end{equation}and
\begin{eqnarray}
R_{k\gamma,l\lambda,i\alpha,j\beta}&=&(\delta_{\alpha\gamma}\delta_{\beta\lambda}
-\delta_{\alpha\lambda}\delta_{\beta\gamma})c_{\alpha\beta}\partial_i\partial_j\partial_k\partial_lG\nonumber\\
&&{}-\delta_{\alpha\gamma}e_{\alpha\mu}\partial_i\partial_k\partial_mG\,
g^{m\mu,n\nu}\delta_{\beta\lambda}e_{\beta\nu}\partial_j\partial_l\partial_nG
\nonumber\\
&&{}+\delta_{\beta\gamma}e_{\beta\mu}\partial_j\partial_k\partial_mG\,
g^{m\mu,n\nu}\delta_{\alpha\lambda}e_{\alpha\nu}\partial_i\partial_l\partial_nG
\end{eqnarray}
where $\alpha$ and $\beta$ are not summed, $Z_{ijkl}=\partial_i\partial_j\partial_k\partial_lG|_{r=0}$ and $\delta_{\alpha\beta\gamma\delta}=1$ if $\alpha=\beta=\gamma=\delta$ and zero otherwise.
The Riemannian curvature tensor can be decomposed into
 \begin{equation}
R_{k\gamma,l\lambda,i\alpha,j\beta}=e_{\alpha\beta}e_{\gamma\lambda}\,S_{ijkl}
-\delta_{\alpha\beta\gamma\lambda}e_{ij}e_{kl}\,\omega+c_{\alpha\beta}c_{\gamma\lambda}e_{ij}e_{kl}\Theta/2,
\end{equation}
where we have used $(\delta_{\alpha\gamma}\delta_{\beta\lambda}+\delta_{\beta\gamma}\delta_{\alpha\lambda})c_{\alpha\beta}=c_{\alpha\beta}c_{\gamma\lambda}$ and 
$(\delta_{\alpha\gamma}\delta_{\beta\lambda}
-\delta_{\beta\gamma}\delta_{\alpha\lambda})c_{\alpha\beta}=e_{\alpha\beta}e_{\gamma\lambda}$ (no summation over $\alpha, \beta$) and the definitions
\begin{eqnarray}
S_{ijkl}&=&\partial_i\partial_j\partial_k\partial_lG+\left(\partial_i\partial_k\partial_pG\partial_j\partial_l\partial_qG+
\partial_j\partial_k\partial_pG\partial_i\partial_l\partial_qG\right)\xi_{pq}/2\nonumber\\
\omega&=&e_{ij}e_{kl}\partial_i\partial_k\partial_mG\partial_j\partial_l\partial_nG\,h^{mn}/2\nonumber\\
\Theta&=&e_{ij}e_{kl}\partial_i\partial_k\partial_mG\partial_j\partial_l\partial_nG\,\xi_{mn}/2.
\end{eqnarray}
We obtain easily
\begin{eqnarray}
g^{k\gamma,i\alpha}R_{k\gamma,l\lambda,i\alpha,j\beta}&=&\left(\delta_{\beta\lambda}h^{ik}-c_{\beta\lambda}\xi_{ik}\right)S_{ijkl}+\delta_{\beta\lambda}e_{ij}e_{kl}h^{ik}(-\omega+\Theta/2)\nonumber\\
&&{}+c_{\beta\lambda}e_{ij}e_{kl}\xi_{ik}\Theta/2
\end{eqnarray}
and
\begin{eqnarray}
R&=&g^{k\gamma,i\alpha}g^{l\lambda,j\beta}R_{k\gamma,l\lambda,i\alpha,j\beta}\nonumber\\
&=&2\left(\left(h^{ik}h^{jl}-\xi_{ik}\xi_{jl}\right)S_{ijkl}-\frac{2\omega}{\det h}+
\frac{\Theta}{\det h}+\Theta\det\xi\right).
\end{eqnarray}
To simplify this expression, we define
\begin{equation}
S_{ijkl}=T_{ijkl}-(2\delta_{ij}\delta_{kl}-\delta_{ik}\delta_{jl}-\delta_{il}\delta_{jk})\Theta/2,
\end{equation}
where $T_{ijkl}$ can be written as
\begin{equation}
T_{ijkl}=\partial_i\partial_j\partial_k\partial_lG+\partial_i\partial_j\partial_pG\partial_k\partial_l\partial_qG\xi_{pq}.
\end{equation} The scalar curvature is now
\begin{equation}
R= 2\left(\left(h^{ik}h^{jl}-\xi_{ik}\xi_{jl}\right)T_{ijkl}-\frac{2\omega}{\det h}+
\frac{2\Theta}{\det h}\right).
\end{equation}
Furthermore
\begin{equation}
\fl h^{ik}h^{jl}\partial_i\partial_j\partial_mG\partial_k\partial_l\partial_nG\xi_{mn}+2\Theta/\det h=h^{ij}\partial_i\partial_j\partial_mGh^{kl}\partial_k\partial_l\partial_nG\xi_{mn}.
\end{equation}
For convenience we choose $\bi{r}^A-\bi{r}^B=(r,0)$.
The non-vanishing partial derivatives of $G(r)$ for that particular choice of $\bi{r}$ are listed in equation (\ref{partial}). We have $\omega=Z_1'Z_2'/D_1-(Z_2')^2/D_2$ and
\begin{equation}
\label{omega}
\fl r\frac{\omega}{\sqrt{\det h}}=\frac{1}{\sqrt{D_1D_2}}\left(
\frac{Z_1'(Z_1-Z_2)}{D_1}-\frac{Z_2'(Z_1-Z_2)}{D_2}\right)=\left(\frac{D}{\sqrt{D_1D_2}}\right)'
\end{equation}
where $D=1-Z_1Z_2$.
The integral $\int\rmd^2r(\det h)^{-1/2}\omega$ is therefore zero up to boundary contributions from $r=0$ and $r\rightarrow\infty$.
The integral over the scalar curvature $\mathcal{H}=(4\pi)^{-1}\int\rmd^2r\sqrt{\det g}R$ becomes (up to boundary contributions)
\begin{eqnarray}
\fl \mathcal{H}&=&(2\pi)^{-1}\int\rmd^2r\sqrt{\det h}\left(\left(h^{ij}h^{kl}-\xi_{ik}\xi_{jl}\right)T_{ijkl}\right)\nonumber\\
&=& \int_0^\infty r\rmd r(D_1D_2)^{1/2}\left(
\frac{Z_1''}{D_1}+\frac{2Z_2''D}{D_1D_2}+\frac{3Z_2'}{rD_2}+\left(\frac{Z_1'}{D_1}+\frac{Z_2'}{D_2}\right)^2\frac{Z_1}{D_1}\right.\nonumber\\
&&\left.{}-(Z_1')^2\left(\frac{Z_1}{D_1}\right)^3-3(Z_2')^2\frac{Z_1Z_2^2}{D_1D_2^2}\right)\nonumber\\
&=& \int_0^\infty r\rmd r(D_1D_2)^{1/2}\left(
\frac{Z_1''}{D_1}+\frac{2Z_2''D}{D_1D_2}+\frac{3Z_2'}{rD_2}+\frac{(Z_1')^2Z_1}{D_1^2}+2\frac{Z_1'Z_2'Z_1}{D_1^2D_2}\right.\nonumber\\
&&\left.{}+\frac{(Z_2')^2Z_1}{D_1D_2}-2(Z_2')^2\frac{Z_1Z_2^2}{D_1D_2^2}\right).
\end{eqnarray}
$Z_1''$ and $Z_2''$ are eliminated with the help of two partial integrations
\begin{eqnarray}
\lefteqn{ \int_0^\infty r\rmd r(D_1D_2)^{1/2}\frac{Z_1''}{D_1}}\nonumber\\&=&\textrm{b.c.}+ \int_0^\infty r\rmd r(D_1D_2)^{1/2}\left(-\frac{Z_1'}{rD_1}+\frac{Z_1'Z_2'Z_2}{D_1D_2}
-\frac{(Z_1')^2Z_1}{D_1^2}\right),\\
\lefteqn{2 \int_0^\infty r\rmd r(D_1D_2)^{1/2}\frac{Z_2''D}{D_1D_2}
}\nonumber\\
&=&\textrm{b.c.}+2 \int_0^\infty r\rmd r(D_1D_2)^{1/2}\left(
-\frac{Z_2'D}{rD_1D_2}+\frac{Z_1'Z_2'Z_2}{D_1D_2}+\frac{(Z_2')^2Z_1}{D_1D_2}\right.\nonumber\\
&&\left.{}-DZ_2'\left(\frac{Z_1'Z_1}{D_1^2D_2}+\frac{Z_2'Z_2}{D_1D_2^2}\right)
\right)
\end{eqnarray}
yielding
\begin{eqnarray}
\mathcal{H}&=&\textrm{b.c.}+ \int_0^\infty r\rmd r(D_1D_2)^{1/2}\left(
-\frac{Z_1'}{rD_1}-\frac{2Z_2'D}{rD_1D_2}+\frac{3Z_2'}{rD_2}\right.\nonumber\\
&&\left.{}+\frac{3Z_1'Z_2'Z_2}{D_1D_2}+
\frac{3(Z_2')^2Z_1}{D_1D_2}-\frac{2(Z_2')^2Z_2}{D_1D_2^2}+
\frac{2Z_1'Z_2'Z_1^2Z_2}{D_1^2D_2}\right)\nonumber\\
&=&\textrm{b.c.}+ \int_0^\infty r\rmd r(D_1D_2)^{1/2}\left(
-\frac{Z_1'}{rD_1}-\frac{2Z_2'D}{rD_1D_2}+\frac{3Z_2'}{rD_2}\right.\nonumber\\
&&\left.{}+\frac{Z_1'Z_2'Z_2}{D_1D_2}+
\frac{3(Z_2')^2Z_1}{D_1D_2}-\frac{2(Z_2')^2Z_2}{D_1D_2^2}+
\frac{2Z_1'Z_2'Z_2}{D_1^2D_2}\right).
\end{eqnarray}
We replace a single $Z_2'=(Z_1-Z_2)/r$ in the last four terms of the integral
\begin{eqnarray}
\fl \mathcal{H}&=&\textrm{b.c.}+ \int\rmd r(D_1D_2)^{1/2}\left(
-\frac{Z_1'}{D_1}-\frac{2Z_2'D}{D_1D_2}+\frac{3Z_2'}{D_2}+\frac{Z_1'(Z_1-Z_2)Z_2}{D_1D_2}\right.\nonumber\\
&&\left.{}+\frac{3Z_2'(Z_1-Z_2)Z_1}{D_1D_2}
-\frac{2Z_2'(Z_1-Z_2)Z_2}{D_1D_2^2}+\frac{2Z_1'(Z_1-Z_2)Z_2}{D_1^2D_2}\right).
\end{eqnarray}
Identity (\ref{omega}) allows us to simplify the last two terms of the above equation
\begin{eqnarray}
2 \int\rmd r\frac{(Z_1-Z_2)Z_2}{\sqrt{D_1D_2}}\left(\frac{Z_1'}{D_1}-\frac{Z_2'}{D_2}\right)&=&2 \int\rmd rZ_2\left(\frac{D}{\sqrt{D_1D_2}}\right)'\nonumber\\
&=&\textrm{b.c.}- \int\rmd r\frac{2Z_2'D}{\sqrt{D_1D_2}}.
\end{eqnarray}
We gather terms $\propto Z_1'$ and $Z_2'$ and gain the remarkably simple result
\begin{equation}
\mathcal{H}=- \int\rmd r\frac{D}{\sqrt{D_1D_2}}(Z_1+Z_2)'
\end{equation}
up to contributions from the boundary.


\section{Functional derivative of the Lagrangian}
\label{sec-lagrangian}
We calculate the variation of
\begin{equation}
\mathcal{L}=-\int\rmd r\left((D_2/D_1)^{1/2}Z_1'+(D_1/D_2)^{1/2}Z_2'\right)
\end{equation}
under an infinitesimal variation of the correlation function $G(r)\rightarrow G(r)+\eta(r)$. We introduce the shorthand notations
$\partial_1=\partial/\partial Z_1$ and $\partial_2=\partial/\partial Z_2$ and obtain
\begin{eqnarray}
\delta\mathcal{L}&=&-\int\rmd r\left[\left((D_2/D_1)^{1/2}\eta'''+(D_1/D_2)^{1/2}(\eta'/r)'\right)\right.\nonumber\\
&&{}+
\left(\partial_1(D_2/D_1)^{1/2}\eta''+\partial_2(D_2/D_1)^{1/2}\eta'/r\right)Z_1'\nonumber\\
&&\left.{}+
\left(\partial_1(D_1/D_2)^{1/2}\eta''+\partial_2(D_1/D_2)^{1/2}\eta'/r\right)Z_2'\right]\\
&=&\textrm{b.c.}+\int\rmd r\left[\left(\partial_1(D_2/D_1)^{1/2}Z_1'+\partial_2(D_2/D_1)^{1/2}Z_2'\right)\eta''\right.\nonumber\\
&&\left.{}+\left(\partial_1(D_1/D_2)^{1/2}Z_1'+\partial_2(D_1/D_2)^{1/2}Z_2'\right)\eta'/r\right]\nonumber\\
&&{}-\int\rmd r\left[\left(\partial_1(D_2/D_1)^{1/2}\eta''+\partial_2(D_2/D_1)^{1/2}\eta'/r\right)Z_1'\right.\nonumber\\
&&\left.{}+
\left(\partial_1(D_1/D_2)^{1/2}\eta''+\partial_2(D_1/D_2)^{1/2}\eta'/r\right)Z_2'\right]\\
&=&\textrm{b.c.}-\int\rmd r\left[\left(\partial_1(D_1/D_2)^{1/2}-\partial_2(D_2/D_1)^{1/2}\right)
\left(Z_2'\eta''-Z_1'\eta'/r\right)\right]\nonumber\\
&=&\textrm{b.c.}+\int\rmd r\,Y\left(Z_2'\eta''-Z_1'\eta'/r\right)
\end{eqnarray}
where
\begin{equation}
Y\equiv\frac{Z_1-Z_2}{(D_1D_2)^{1/2}}.
\end{equation}
The variation becomes
\begin{eqnarray}
\delta\mathcal{L}&=&\textrm{b.c.}-\int\rmd r\,\eta'\left((YZ_2')'+YZ_1'/r\right)\nonumber\\
&=&\textrm{b.c.}+\int\rmd r\frac{\eta'}{r}\left(2Y(Z_1'-Z_2')+Y'(Z_1-Z_2)\right)
\end{eqnarray}or
\begin{equation}
\frac{1}{r}\frac{\delta\mathcal{L}}{\delta G}=\frac{1}{r}\frac{\partial}{\partial_r}\left(\frac{\left(Y(Z_1-Z_2)^2\right)'}{r(Z_1-Z_2)}\right)=\psi'/r=(2\pi)^2C(r).
\end{equation}


\end{document}